\documentclass[AMA]{USG}
\usepackage{anyfontsize}
\usepackage{colortbl}   
\usepackage{steinmetz}
\graphicspath{{./images/}}

\usepackage{amsmath, amsfonts, bm}
\usepackage{adjustbox}
\usepackage{booktabs, multirow}
\usepackage{xcolor}
\usepackage{hyperref, url}

\usepackage{fancyhdr}
\fancypagestyle{plain}{%
  \fancyhf{}%
  \fancyfoot[C]{\thepage}%
}
\fancypagestyle{headings}{%
  \fancyhf{}%
  \fancyfoot[C]{\thepage}%
}
\fancypagestyle{titlepage}{%
  \fancyhf{}%
  \fancyfoot[C]{\thepage}%
}
\DeclareMathOperator{\logit}{logit}
\DeclareMathOperator{\expit}{expit}
\newcommand{\balpha}{\bm{\alpha}}
\newcommand{\bbeta}{\bm{\beta}}
\newcommand{\bdelta}{\bm{\delta}}
\newcommand{\btheta}{\bm{\theta}}
\newcommand{\bomega}{\bm{\omega}}
\newcommand{\hCov}{\widehat{\operatorname{Cov}}}
\newcommand{\bZ}{\bm{Z}}

\begin{document}

\title{Two-Phase Sampling Designs and Analysis Approaches for Ordinal Outcomes}

\author[1]{Yunbi Nam}
\author[2]{Nathan I. Shapiro}
\author[3]{Eric P. Schmidt}
\author[4]{Wesley H. Self}
\author[1,5,*]{Ran Tao}
\author[1,*]{Jonathan S. Schildcrout}

\address[1]{Department of Biostatistics, Vanderbilt University Medical Center, Nashville, Tennessee, USA}
\address[2]{Department of Emergency Medicine, Beth Israel Deaconess Medical Center and Harvard Medical School, Boston, Massachusetts, USA}
\address[3]{Department of Medicine, Mass General Brigham, Boston, Massachusetts, USA}
\address[4]{Vanderbilt Institute for Clinical and Translational Research and Department of Emergency Medicine, Vanderbilt University Medical Center, Nashville, Tennessee, USA}
\address[5]{Vanderbilt Genetics Institute, Vanderbilt University Medical Center, Nashville, Tennessee, USA}
\address[*]{These authors contributed equally to this work.}

\corres{Yunbi Nam (\email{yunbi.nam@vanderbilt.edu})}

\fundingInfo{National Institute of Health, Grant Number: R01HL094786 and R01AI131771}


\abstract[ABSTRACT]{Modern clinical trials and cohort studies gather low-cost data on all participants but may have limited resources to assess expensive exposures such as biomarkers or genomic data. When interest lies in associations involving expensive exposures, two-phase designs provide a cost-effective framework by using information available on all participants to guide the targeted selection of a subset for additional measurements. We extend this framework to studies with ordinal outcomes, a common yet previously unexplored setting. We propose three outcome-informed phase 2 sampling designs---outcome-dependent sampling (ODS), covariate-stratified ODS, and residual-dependent sampling---that leverage phase 1 data to enrich phase 2 selection with informative subjects. We then develop analysis methods for valid and efficient estimation/inference, including conditional likelihood methods with ascertainment-corrected maximum likelihood estimation, multiple imputation, and a full likelihood method using sieve maximum likelihood estimation. Across a range of scenarios, simulation studies show that the proposed methods substantially improve efficiency over simple random sampling with standard maximum likelihood estimation. We further demonstrate their practical utility by examining the association between interleukin-6 and a four-level clinical status outcome---discharged, hospitalized but not in the ICU, hospitalized in the ICU, and death---14 days after randomization into the Crystalloid Liberal or Vasopressors Early Resuscitation in Sepsis trial.}

\contributed{Ran Tao and Jonathan S. Schildcrout contributed equally to this work.}

\maketitle
\setlength{\parskip}{1em}
\section{Introduction} \label{sec:intro}

Modern clinical trials and cohort studies routinely collect low-cost, essential data (\textit{e.g.}, treatment, baseline covariates, and outcomes) on all participants and collect and store biospecimens for future use. Biospecimen storage creates unique opportunities to address new scientific questions, after the primary study has been completed, about biomarker associations with outcomes or treatment effectiveness. However, biomarker measurement from stored samples is often expensive and therefore infeasible for the full cohort. In these settings, selecting a subset of participants for measurement is needed. Two-phase study designs provide a cost-effective framework for deciding which participants to select for such measurements: the selection of the phase 2 subset is informed by readily available phase 1 data, allowing for the strategic allocation of limited resources to measure expensive exposures \cite{Neyman1938, White1982}. The success of this framework hinges on two key elements: judicious sampling of participants for exposure ascertainment and valid, efficient statistical analysis of collected data.

Enriching phase 2 samples with informative subjects---typically those with more extreme outcome values---has been shown to improve statistical efficiency. For binary outcomes, case-control designs, wherein cases are oversampled, are particularly effective for rare diseases \cite{Breslow1996}, while for continuous outcomes, oversampling individuals in the tails improves efficiency relative to simple random sampling (SRS) \cite{Zhou2002, Lin2013}. Efficiency can be further increased by stratifying on both the outcome and previously collected, phase 1 covariates \cite{breslow1999}. More recent research has demonstrated efficiency gains by oversampling individuals with extreme residuals from a working regression model of the outcome on inexpensive covariates \cite{Lin2013, sun2017, tao2020, digravio2024}.

While efficient sampling designs can yield highly informative phase 2 subcohorts, it is well known that naive analyses (\textit{i.e.}, those that ignore the sampling scheme) generally lead to biased estimates. One well-known exception involves logistic regression under case-control designs where valid inferences regarding covariate-outcome associations are obtained even when the biased sampling study design is ignored \cite{holt1980, prentice1979}. Beyond such special cases, valid inference requires accounting for the sampling design, either through design-based or likelihood-based methods. Design-based methods represent the full phase 1 cohort by weighting the phase 2 subcohort using known selection probabilities, for example, through inverse probability weighting \cite{horvitz1952, deville1993}, while likelihood-based methods explicitly incorporate the selection mechanism into a model-based likelihood. While design-based methods offer robustness to model misspecification, likelihood-based methods can be substantially more efficient than design-based methods \cite{amorim2021}. In this paper, we focus on developing valid and efficient likelihood-based inference procedures.

Within likelihood-based methods, we broadly classify analysis approaches into complete-case and full-data approaches. Complete-case approaches use only data collected from subjects selected in phase 2 and correct for biased sampling by incorporating the sampling probabilities into a conditional likelihood \cite{wild1991, schildcrout2008}. In contrast, full-data approaches retain the entire phase 1 cohort by combining complete data from subjects selected for phase 2 with partial data (\textit{i.e.}, covariates and outcomes but not the expensive exposure) from those not selected. For example, Schildcrout et al. \cite{schildcrout2015} proposed multiple imputation with a parametric imputation model to fill in unobserved exposure in unselected subjects, while Tao, Zeng, and Lin \cite{tao2017} developed a semiparametric likelihood method that incorporates subjects with missing exposures by nonparametrically approximating the conditional exposure distribution.

While most research on two-phase designs and analyses have focused on continuous, binary, and time-to-event outcomes, clinical and epidemiological studies often collect ordinal outcomes to capture graded disease severity and clinical progression. The use of such ordinal endpoints has grown in recent years, particularly during the COVID-19 pandemic when the World Health Organization introduced an ordinal clinical progression scale to standardize COVID-19 severity assessment across trials \cite{who2020, self2021}. Ordinal endpoints also arise in other clinical settings; for example, the Crystalloid Liberal or Vasopressors Early Resuscitation in Sepsis (CLOVERS) trial collected daily patient status data, which allowed construction of many ordinal outcomes including hospital-free days and ICU-free days \cite{clovers2023}. Despite the increasing prevalence of ordinal outcomes in clinical research, specifically in critical care research, two-phase methods have not yet been extended to accommodate ordinal outcomes.

To address this gap, we develop the first unified framework for two-phase studies with ordinal outcomes, making three distinct contributions. First, we derive efficient phase 2 sampling designs specifically tailored to ordinal outcomes under the proportional odds model, including outcome-dependent sampling (ODS), covariate-stratified ODS (CSODS), and residual-dependent sampling (RDS). These designs exploit the ordered structure of the outcome to concentrate sampling resources on the most informative subjects---a strategy that, to our knowledge, has not previously been studied for ordinal data. Second, we develop three complementary likelihood-based inference procedures that are valid under these designs: a complete-case approach based on ascertainment-corrected maximum likelihood (ACML), and two full-data approaches---multiple imputation (MI) and sieve maximum likelihood estimation (SMLE)---that recover information from unsampled subjects. Third, we evaluate the finite-sample operating characteristics of all proposed design-analysis combinations in the ordinal outcome setting through extensive simulations and a plasmode study, demonstrating that the proposed designs and estimators yield substantial efficiency gains over simple random sampling with standard analyses.

The rest of the paper is organized as follows. Section 2 introduces the proportional odds model and establishes notation. Section 3 develops the phase 2 sampling designs, followed by inference procedures in Section 4. Section 5 evaluates the finite-sample operating characteristics of the proposed design–analysis approaches through simulation studies. Section 6 demonstrates their practical utility in a plasmode simulation study \cite{franklin2014} using data from the CLOVERS trial. Finally, Section 7 concludes with a discussion.

\section{The Proportional Odds Model} \label{sec:model}

Consider an ordinal outcome $Y \in \{1, \dots, k\}$ with $k$ ordered categories, an expensive exposure $X$ of primary interest, and a vector of inexpensive covariates $\bZ$. We adopt the proportional odds (PO) model \cite{mccullagh1980}, a widely used framework for analyzing ordinal outcomes, and briefly outline its formulation below:
\begin{equation} \label{eq:po}
    \logit P(Y \leq j \mid X, \bZ) 
    = \alpha_j + \beta_x\, X + \bbeta_z^{\top}\, \bZ, 
    \quad j = 1, \dots, k - 1,
\end{equation}
where $\alpha_1 < \alpha_2 < \cdots <\alpha_{k-1}$ are category-specific intercepts. This model characterizes cumulative probabilities, assuming that covariate effects are consistent across all $j = 1, \ldots, k - 1$. Our primary interest lies in estimating the parameter $\beta_x$, which measures the log-odds ratio associated with a unit change in exposure $X$.

Let $N$ denote the total number of subjects in the study cohort. Let $\balpha = (\alpha_1, \alpha_2, \dots, \alpha_{k-1})^{\top}$ and $\btheta = (\balpha^{\top}, \beta_x, \bbeta_z^{\top})^{\top}$ be the regression parameters. For subject $i \in \{1, \dots, N\}$, the probability mass function is given by:
\begin{equation*}
    P(Y_i = j \mid  X_i, \bZ_i) = 
    \begin{cases}
    \expit(\alpha_1 + \beta_x\, X_i + \bbeta_z^{\top}\, \bZ_i) & \text{if } j = 1 \\
    \expit(\alpha_j + \beta_x\, X_i + \bbeta_z^{\top}\, \bZ_i) - \expit(\alpha_{j-1} + \beta_x\, X_i + \bbeta_z^{\top}\, \bZ_i) & \text{if } 2 \leq j \leq k-1 \\
    1 - \expit(\alpha_{k-1} + \beta_x\, X_i + \bbeta_z^{\top}\, \bZ_i) & \text{if } j = k
    \end{cases}
\end{equation*}
where $\expit(x) = \{1 + e^{-x}\}^{-1}$ is the inverse-logit function. Under the assumption that observations are independent and the PO model in equation~(\ref{eq:po}) is correctly specified, the maximum likelihood estimator (MLE) of $\btheta$ can be obtained by maximizing the log-likelihood function:
\begin{equation*}
    l_N(\btheta) = \sum_{i=1}^N \log P(Y_i \mid  X_i, \bZ_i; \btheta).
\end{equation*}

\section{Two-Phase Sampling Designs} \label{sec:design}

The overarching strategy of efficient two-phase study designs is to increase observed response variability compared to simpler SRS designs so that outcome-exposure associations can be detected. ODS achieves this by oversampling subjects with extreme values of $Y$. However, when a baseline covariate or confounder $Z_1 \in \bZ$ is strongly associated with $Y$, sampling on $Y$ alone may not sufficiently increase variability in $Y$ conditional on $Z_1$, yielding limited efficiency gains for estimating the $X$-$Y$ association. CSODS addresses this limitation by defining sampling strata jointly on $(Y, Z_1)$, explicitly increasing variability in $Y$ within levels of $Z_1$. When multiple inexpensive covariates are correlated with $Y$, further gains may be achieved by targeting variability in the residual component of $Y$ after accounting for $\bZ$. RDS implements the idea by prioritizing subjects whose outcomes are poorly explained by a working model that regresses $Y$ on $\bZ$ (\textit{i.e.}, those with relatively large residuals in absolute value). 

\subsection{Outcome-Dependent Sampling Design}

Let $S$ denote the selection indicator for phase 2, with $S_i = 1$ if subject $i$ is selected for measurement of $X_i$ and $S_i = 0$ if not. As in case-control designs for binary outcomes, the fixed number of categories in an ordinal outcome makes it straightforward to define sampling strata for the ODS design. For an ordinal outcome with $k$ categories, the sampling probability depends exclusively on the observed outcome level $j \in \{1, \dots, k\}$ and is defined as $\pi(j) = P(S = 1 \mid Y = j)$. The goal of targeted outcome-dependent sampling is to increase variability in $Y$ in phase 2. This potentially allows for more efficient estimation of the exposure-outcome association compared to standard SRS designs.

\subsection{Covariate-Stratified Outcome-Dependent Sampling Design}

Let the sampling strata be indexed by $(j, g)$, where $j \in \{1, \dots, k\}$ denotes the outcome category and $g \in \{1, \dots, t\}$ indexes groups of covariate $Z_1$ defined by a stratification function $G(\cdot)$. For categorical $Z_1$, these groups may correspond to the observed levels, whereas for continuous $Z_1$, $G(Z_1)$ represents intervals obtained by discretization (\textit{e.g.}, based on quantiles or clinically meaningful thresholds). When $t$ is large, the resulting $k \times t$ stratification can yield many sparsely populated cells, making it difficult to specify reliable sampling probabilities within each cell. One may then collapse adjacent categories of categorical $Z_1$ or discretize continuous $Z_1$ into a manageable number of strata. The sampling probability is then $\pi(j, g) = P(S = 1 \mid Y = j, G(Z_{1}) = g)$. The joint stratification on $(Y, Z_1)$ is designed to increase the variability of $Y$ within levels of $Z_1$ in phase 2, thereby improving efficiency for estimating the exposure-outcome association relative to SRS.

\subsection{Residual-Dependent Sampling Design}

To conduct an RDS design, we fit the PO model using phase 1 data:
\begin{equation}
    \logit P(Y \leq j \mid \bZ) 
    = \alpha_j^* + \bbeta_z^{*\top}\, \bZ, 
    \quad j = 1, \dots, k - 1,
\end{equation}
and compute a subject-specific residual quantity for each subject $i$. Phase 2 sampling can then be guided toward increasing variability in the residual distribution, thereby enriching the phase 2 sample with informative individuals for efficient estimation of the $X$-$Y$ association.

While other papers have shown that, with other outcome distributions, RDS procedure can be highly efficient \cite{sun2017, digravio2024}, defining a single residual per subject under the PO model is not straightforward, as multiple intercepts induce multiple, cutpoint-specific residuals. To address this challenge, we appeal to the probability-scale residual (PSR) \cite{li2012}. The PSR has two features that are important for RDS designs: (i) it yields a single residual per subject; and (ii) it captures both the direction and monotonic magnitude of deviation between observed and fitted outcomes. 

Formally, let $Y$ be an ordinal variable with true distribution $F$, $F^*$ be the fitted distribution, and $Y^*$ be a random variable drawn from $F^*$. For an observed outcome $y$, the PSR is defined as $r(y, F^*) = E\{\text{sign}(y, Y^*)\} = P(Y^* < y) - P(Y^* > y)$, where $\text{sign}(a, b)$ is $-1$ if $a < b$, $0$ if $a = b$, and $1$ if $a > b$. By construction, PSR takes values in $[-1, 1]$ and measures the discrepancy between observed and fitted outcomes on a common probability scale.

\section{Analysis Approaches} \label{sec:analysis}

Under the sampling designs described in Section~\ref{sec:design}, maximum likelihood estimation can lead to biased estimates if the sampling mechanism is ignored \cite{holt1980}. We therefore develop three likelihood-based analysis approaches that properly account for the sampling design. First, ACML is a complete-case approach that restricts analysis to phase 2 subjects with $X$ observed while acknowledging the sampling scheme in the likelihood. The other two methods---MI and SMLE---retain the full phase 1 cohort and incorporate the partial information---$(Y, \bZ)$ but not $X$---for those subjects not sampled for phase 2. MI does so by treating unmeasured $X$ as missing and then imputing its values from a parametric exposure model for subjects not selected in phase 2, whereas SMLE estimates the distribution of $X$ nonparametrically with B-spline sieves to improve robustness against exposure-model misspecification and then integrates over $X$ for subjects not selected in phase 2.

\subsection{Ascertainment Corrected Maximum Likelihood}

 We adopt a conditional likelihood framework to the setting of an ordinal outcome under the PO model. In this framework, inference is based on the conditional distribution of subjects selected into phase 2. Specifically, the ascertainment-corrected log-likelihood for the $n$ sampled subjects is given by:
\begin{equation} \label{eq:acll}
\begin{split}
    l_n^c(\btheta)
    &= \sum_{i=1}^{n} \log P(Y_i \mid X_{i}, \bZ_{i}, S_i = 1; \btheta) \\
    &\propto \sum_{i=1}^{n} \Bigg(\log P_{\btheta}(Y_i \mid X_i, \bZ_i)- \log \underbrace{\left\{ \sum_{j=1}^k \pi(S_i=1 \mid Y_i=j, \bZ_i)\, P_{\btheta}(Y_i=j \mid X_i, \bZ_i)\right\}}_{\text{AC}_i}\Bigg),
\end{split}
\end{equation}
where $\text{AC}_i$ is the ascertainment-correction (AC) term that adjusts the likelihood contribution of subject $i$ due to the biased phase 2 sampling design.

In ODS and CSODS designs, these sampling probabilities are explicitly specified by design, making the AC component straightforward to construct. In contrast, RDS selects subjects based on probability-scale residuals rather than strata defined by $(Y, \bZ)$, so $\pi(S_i = 1 \mid Y_i, \bZ_i)$ cannot be expressed in a simple closed form. We therefore restrict ACML to the ODS and CSODS settings.

Estimates of $\btheta$ and its covariance matrix are derived by maximizing the ascertainment-corrected log-likelihood in equation~(\ref{eq:acll}) using the Newton-Raphson algorithm. The corresponding score function is obtained by subtracting the derivative of $\log(\text{AC}_i)$ from the derivative of the standard log-likelihood contribution under the PO model. The limiting covariance matrix of $\hat{\btheta}$ is then estimated using the inverse of the observed information matrix.

\subsection{Multiple Imputation}

In two-phase studies, because $X$ is ascertained in participants sampled in phase 2 and it is missing in participants not sampled, the resulting data structure can be naturally framed within a missing-data framework. Unlike general missing-data settings where the missingness mechanism is unknown, here it is known by design because it is fully determined by the observed phase 1 data. Thus, the missing at random (MAR) assumption holds by construction:
\begin{equation} \label{eq:mar}
    P(X_i \mid Y_i, \bZ_i, S_i = 0) 
    \stackrel{\text{(a)}}{=} P(X_i \mid Y_i, \bZ_i)
    = P(X_i \mid Y_i, \bZ_i, S_i = 1).
\end{equation}
Under this framework, we can impute $X$ values for individuals with $S_i = 0$, using data from those with $S_i = 1$, and in doing so, we preserve information contained in $(Y, \bZ)$ that would otherwise be lost in a complete-case analysis.

Schildcrout et al. (2015) describes two approaches for constructing the imputation model. In this work, we build upon the approach that combines the response model $[Y_i \mid X_i, \bZ_i]$ with the exposure model $[X_i \mid \bZ_i]$. Specifically, using Bayes' Theorem and equality (a) in equation~(\ref{eq:mar}), our imputation model is given by:
\begin{equation} \label{eq:exposure_mi}
    P(X_i \mid Y_i, \bZ_i) 
    = \frac{P(Y_i \mid X_i, \bZ_i)\, P(X_i \mid \bZ_i)}{P(Y_i \mid \bZ_i)}
    = \frac{P(Y_i \mid X_i, \bZ_i)\, P(X_i \mid \bZ_i)}{\int_{\mathcal{X}} P(Y_i \mid X_i = x, \bZ_i)\, dF(x \mid \bZ_i)},
\end{equation}
where $\mathcal{X}$ denotes the support of $X$ and $F(\cdot \mid \bZ_i)$ is the conditional distribution function of $X$. The denominator marginalizes over $X$: when $X$ is discrete, this reduces to a sum over $\mathcal{X}$; when $X$ is continuous, it becomes $\int_{\mathcal{X}} P(Y_i \mid X_i = x, \bZ_i)\, f(x \mid \bZ_i)\, dx$. In practice, for continuous $X$, we approximate this integral by summing over a grid of $C$ equally spaced candidate values spanning a plausible range (\textit{e.g.}, the range of observed $X$ values), with $C$ chosen sufficiently large to ensure accurate approximation.

The MI procedure with $M$ imputations proceeds as follows:
\begin{enumerate}
    \item For the subjects selected into phase 2 ($S_i = 1$), fit the proportional odds model $[Y_i \mid X_i, \bZ_i; \bdelta]$ using ACML to obtain $\hat{\bdelta}$ and $\hCov(\hat{\bdelta})$. 
    \item For the subjects selected into phase 2 ($S_i = 1$), fit the regression model $[X_i \mid \bZ_i; \bomega]$ using inverse probability weighting (IPW) to obtain $\hat{\bomega}$ and $\hCov(\hat{\bomega})$.
    \item Construct a grid of $C$ equally spaced candidate values $\{x_1, \dots, x_C\}$ spanning the observed range of $X$.
    \item For each imputation $m=1, \dots, M$,
    \begin{enumerate}[a.]
        \item Draw $\bdelta^{(m)} \sim N(\hat{\bdelta}, \hCov(\hat{\bdelta}))$ and $\bomega^{(m)} \sim N(\hat{\bomega}, \hCov(\hat{\bomega}))$.
        \item For each subject $i$ with $S_i = 0$,
        \begin{enumerate}[i.]
            \item For $c = 1, \dots, C$, compute:
            $$
            w_c^{(m)} = P(Y_i \mid X_i = x_c, \bZ_i; \bdelta^{(m)})\, P(X_i = x_c \mid \bZ_i; \bomega^{(m)}) 
            $$
            \item Normalize weights:
            $$
            \tilde{w}_c^{(m)} = \frac{w_c^{(m)}}{\sum_{c'=1}^C w_{c'}^{(m)}}.
            $$
            \item Impute $X_i$ by drawing from $\{x_1, \ldots, x_C\}$ with probabilities $\{\tilde{w}_1^{(m)}, \dots, \tilde{w}_C^{(m)}\}$.
        \end{enumerate}
        \item Fit the proportional odds model $[Y_i \mid X_i, \bZ_i; \btheta]$ on the $m$th imputed dataset to obtain $\hat{\btheta}^{(m)}$ and $\hCov(\hat{\btheta}^{(m)})$.
    \end{enumerate}
    \item Combine the $M$ sets of estimates using Rubin's rules.
\end{enumerate}

Because step (1) relies on ACML for fitting the response model, this MI procedure is readily applicable under the ODS and CSODS designs, but not under the RDS design. For the number of imputations $M$, we follow the practical rule of thumb suggested by White, Royston, and Wood \cite{white2011}, which recommends using $M \geq 100 \times \text{FMI}$, where FMI is the fraction of missing information, to achieve adequate reproducibility of key inferential results.

\subsection{Sieve Maximum Likelihood Estimation}

Tao et al. \cite{tao2017} proposed efficient semiparametric inference method for general two-phase studies, referred to as the sieve maximum likelihood estimation (SMLE). We extend this framework to accommodate ordinal outcomes. Let inexpensive covariates be partitioned as $\bZ = (\bZ_1, \bZ_2)$, where $\bZ_1$ consists of covariates potentially correlated with $X$, and $\bZ_2$ consists of those assumed to be independent of $X \mid \bZ_1$. The joint distribution of $(Y, X, \bZ)$ can be factorized as $P(Y \mid X, \bZ; \btheta)\, H(X \mid \bZ_1)\, P(\bZ_1, \bZ_2)$. The MAR assumption in equation~(\ref{eq:mar}) implies that the phase 2 sampling indicator $S$ depends on $(Y, X, \bZ)$ only through the phase 1 data $(Y, \bZ)$. This allows us to ignore $S$ in the full likelihood for estimation of $\btheta$. The observed-data log-likelihood is then given by:
\begin{equation} \label{eq:obs_ll_sml}
\begin{split}
    l_N^f(\btheta, H) =
    &\sum_{i=1}^N S_i\big\{\log P(Y_i \mid X_i, \bZ_i; \btheta) + \log H(X_i \mid \bZ_{1i})\big\} \\
    &+ \sum_{i=1}^N (1 - S_i) \log \int_{\mathcal{X}} P(Y_i \mid x, \bZ_i; \btheta)\, H(x \mid \bZ_{1i})\, dx.
\end{split}
\end{equation}
This semiparametric formulation combines a parametric model for $P(Y \mid X, \bZ; \btheta)$ with a nonparametric estimation of $H(X \mid \bZ_1)$, potentially improving robustness against misspecification of the exposure model. 

Let $d (\leq n)$ denote the total number of distinct observed values of $X$, taking values in $\{x_1, \dots, x_d\}$. For each observed value of $\bZ_1$, the conditional distribution $H(X \mid \bZ_1)$ can be estimated nonparametrically as a discrete probability distribution supported on $\{x_1, \dots, x_d\}$. However, when only a few observations of $X$ correspond to each value of $\bZ_1$---such as when $\bZ_1$ includes continuous variables---this approach is not feasible. To overcome this, SMLE applies the method of sieves with B-spline basis functions to approximate $H$ \cite{tao2017}. Assuming $\bZ_1$ has bounded support, we define a set of B-spline basis functions $\{B_l^q(\bZ_{1}): l = 1, \ldots, s_n\}$, where $B_l^q$ is the $l$th B-spline function of degree $q$, and $s_n$ denotes the number of basis functions. The approximating functions for estimating $H$ in equation~(\ref{eq:obs_ll_sml}) are given by:
\begin{equation} \label{eq:Gapprox}
    \begin{split}
        \log H(X_i \mid \bZ_{1i}) 
        &\approx \sum_{v=1}^d I(X_i = x_v)\, \sum_{l=1}^{s_n} B_l^q(\bZ_{1i})\, \log p_{vl}, \\
        H(X_i \mid \bZ_{1i}) 
        &\approx \sum_{v=1}^d I(X_i = x_v)\, \sum_{l=1}^{s_n} B_l^q(\bZ_{1i})\, p_{vl}, 
    \end{split}
\end{equation}
where $p_{vl}$ is the coefficient corresponding to the $l$th basis function at $x_v$, capturing its contribution to $H(X_i=x_v \mid \bZ_{1i})$. We can estimate $\btheta$ and $\{p_{vl}\}$ by maximizing the following function:
\begin{equation} \label{eq:obs_ll2_sml}
\begin{adjustbox}{max width=\linewidth}
$
    \begin{aligned}
        l_N^f(\btheta, \{p_{vl}\}) = 
        &\sum_{i=1}^N S_i \left\{\log P(Y_i \mid X_i, \bZ_i; \btheta) + \sum_{v=1}^d I(X_i = x_v)\, \sum_{l=1}^{s_n} B_l^q(\bZ_{1i})\, \log p_{vl}\right\} \\
        &+ \sum_{i=1}^N (1 - S_i)\, \log 
        \left\{
        \sum_{v=1}^d  P (Y_i \mid x_v, \bZ_i; \btheta)\, \sum_{l=1}^{s_n} B_l^q(\bZ_{1i})\, p_{vl}
        \right\}
    \end{aligned}
$
\end{adjustbox}
\end{equation}
under the constraints of $p_{vl} \geq 0$ for all $v = 1, \dots, d$ and $l = 1, \dots, s_n$, and $\sum_{v=1}^d p_{vl} = 1$ for each $l = 1, \dots, s_n$.

The Appendix provides details of the expectation-maximization (EM) algorithm used to obtain the maximum likelihood estimates $\btheta$ from expression~(\ref{eq:obs_ll2_sml}). Asymptotic properties of the resulting estimator $\hat{\btheta}$---consistency, asymptotic normality, and asymptotic efficiency---were established by Tao et al. \cite{tao2017} in Section S.1 of the Supplementary Material.

To derive covariance estimates for $\hat{\btheta}$, we use the profile log-likelihood function defined by $pl(\btheta) \equiv \max_{\{p_{vl}\}} l_N(\btheta, \{p_{vl}\})$. The profile log-likelihood $pl(\btheta)$ is obtained by holding $\btheta$ fixed in the EM algorithm and computing equation~(\ref{eq:obs_ll2_sml}) at the of convergence the EM algorithm. The asymptotic covariance matrix of $\hat{\btheta}$ is then given by the negative inverse of the Hessian of $pl(\hat{\btheta})$, which is evaluated numerically using a second-order finite difference scheme. The $(v, l)$th element of the Hessian is expressed as $h_N^{-2}\, \{pl(\hat{\btheta} + \bm{e}_v\, h_N + \bm{e}_l\, h_N) - pl(\hat{\btheta} + \bm{e}_v\, h_N) - pl(\hat{\btheta} + \bm{e}_l\, h_N) + pl(\hat{\btheta})\}$, where $\bm{e}_k$ denotes the $v$th canonical vector and $h_N$ is a perturbation constant of order $N^{-1/2}$.

\section{Simulation Studies} \label{sec:sim}
We conducted extensive simulation studies to evaluate the validity and efficiency of the proposed designs and analysis approaches for two-phase studies with an ordinal outcome, as summarized in Table~\ref{tab:simMethods}. For each simulation replicate, we generated a phase 1 cohort of $N = 1{,}500$ subjects with a continuous expensive exposure $X$ and two inexpensive covariates, $Z_1$ and $Z_2$. The variables $(X, Z_1, Z_2)$ were jointly generated from a multivariate normal distribution with mean zero and unit variances, where the correlation between $X$ and $Z_1$ was $0.2$, and $Z_2$ was independent of both. A four-level ordinal outcome $Y$ was then generated under the following PO model:
\begin{equation} 
\logit \left\{ P(Y \leq j \mid X, Z_1, Z_2) \right\} = \alpha_j + \beta_x\, X + \beta_{z1}\, Z_1 + \beta_{z2}\, Z_2, \quad j = 1,2,3.
\end{equation}
The simulation scenarios were designed to  evaluate how the performance of the proposed designs and analysis approaches changes with the number of covariates that are strongly associated with the outcome. Specifically, we considered three scenarios for the regression coefficients $(\beta_x, \beta_{z1}, \beta_{z2})$: (i) $(0.5, -0.1, -0.1)$, (ii) $(0.5, -1.5, -0.1)$, and (iii) $(0.5, -1.5, -1.5)$. The exposure effect $\beta_x$ was fixed at 0.5, while the effect sizes of $\beta_{z1}$ and $\beta_{z2}$ were varied to represent settings with (i) no strongly associated covariate, (ii) one strongly associated covariate, and (iii) two strongly associated covariates. The intercepts $\bm{\alpha} = (\alpha_1, \alpha_2, \alpha_3)$ were calibrated to yield average prevalences of approximately $75\%$, $10\%$, $5\%$, and $10\%$ for outcome levels $1$ through $4$, respectively. 
\begin{center}
\begin{table*}[!ht]
\caption{Combinations of sampling designs (row) and analysis procedures (column) considered in the simulation study. The combination SRS + ML is used as the reference method for comparison. \label{tab:simMethods}}
\begin{tabular*}{\textwidth}{@{\extracolsep\fill}lrrr@{\extracolsep\fill}}
    \toprule
    \textbf{Design} & \textbf{ML/ACML} & \textbf{MI} & \textbf{SMLE} \\
    \midrule
    \textbf{SRS} & SRS + ML & SRS + MI & SRS + SMLE \\
    \textbf{ODS} & ODS + ACML & ODS + MI & ODS + SMLE \\
    \textbf{CSODS} & CSODS + ACML & CSODS + MI & CSODS + SMLE \\
    \textbf{RDS} &  & & RDS + SMLE \\
    \bottomrule
\end{tabular*}
\begin{tablenotes}
\item {\it Abbreviations:} 
SRS = simple random sampling; 
ODS = outcome-dependent sampling; 
CSODS = covariate-stratified outcome-dependent sampling; 
RDS = residual-dependent sampling; 
ML = maximum likelihood estimation; 
ACML = ascertainment-corrected maximum likelihood estimation; 
MI = multiple imputation; 
SMLE = sieve maximum likelihood estimation.
\end{tablenotes}
\end{table*}
\end{center}

In each simulation replicate, we compared three proposed sampling designs with SRS, selecting approximately $n = 400$ subjects from a cohort of $N = 1{,}500$, assuming that $X$ was observed only for those selected into phase 2. The proposed designs shared a common objective of increasing variability in the sampled data, but they differed in how this variability was targeted: ODS targeted variability in $Y$, CSODS targeted variability in $Y$ conditional on $Z_1$, and RDS targeted variability in the residuals from a PO model with $(Z_1, Z_2)$. We found that overly extreme allocations often concentrate sampling in a small number of strata and thus render downstream inference procedures unstable. To avoid overly extreme allocations, a pre-allocation step was incorporated in each targeted design: a small number of subjects was first selected from each stratum (defined by $Y$ in ODS and RDS, and by $(Y, Z_1)$ in CSODS), after which the remaining sample was allocated to maximize variability. 

The phase 2 sampling procedures used in the simulation were as follows. Under SRS, each phase 1 subject was independently selected with equal probability $\pi = 0.27$, yielding an expected phase 2 sample size of $400$. For ODS, Bernoulli sampling was used within outcome levels, with sampling probabilities set to achieve the desired allocations across categories: $\pi(j) = P(S = 1 \mid Y = j) = n_j / N_j, j = 1, \dots, 4$, where $N_j$ is the number of phase 1 subjects in category $j$, and $n_j$ is expected phase 2 sample size from that category. In our simulations, these probabilities correspond to $\pi(j) = (0.19, 0.13, 0.27, 1)$, yielding average phase 2 allocations $(n_1, n_2, n_3, n_4) = (210, 20, 20, 150)$ from outcome categories with average phase 1 sizes $(N_1, N_2, N_3, N_4) = (1125, 150, 75, 150)$. For CSODS, sixteen strata were defined jointly for outcome levels and quartile-based categories of $Z_1$. Within each stratum $(j,g)$, subjects were independently selected with probability $\pi(j, g) = P(S = 1 \mid Y = j, G(Z_1) = g) = n_{jg} / N_{jg}, j,g = 1, \dots, 4$, where where $N_{jg}$ and $n_{jg}$ are the numbers of phase 1 subjects and expected phase 2 allocation in stratum $(j, g)$, respectively (Table~\ref{tab:simCSODS}). For RDS, we fit a PO model using phase 1 data, $\logit \left\{ P(Y \leq j \mid Z_1, Z_2) \right\} = \alpha_j^* + \beta_{z1}^*\, Z_1 + \beta_{z2}^*\, Z_2, \quad j = 1,2,3$, and computed PSRs. Twenty subjects were first randomly selected from each outcome level (1--4), and then 160 subjects were sampled from both the highest and the lowest PSRs among the remaining individuals.
\begin{center}
\begin{table*}[!ht]%
\caption{Stratum-specific sampling probabilities $n_{jg} / N_{jg}$ $(\pi(j,g))$ used for covariate-stratified outcome-dependent sampling under each scenario in simulation studies. \label{tab:simCSODS}}
\begin{tabular*}{\textwidth}{@{\extracolsep\fill}lrrrr@{\extracolsep\fill}}
    \toprule
    & \multicolumn{4}{c}{$\bm{Y}$} \\
    \cmidrule{2-5}
    $\bm{G(Z_1)}$ & \textbf{1} & \textbf{2} & \textbf{3} & \textbf{4} \\
    \midrule
    \multicolumn{5}{l}{\textbf{Scenario 1}} \\
    $\bm{Q_1}$ & 53/281 (0.19) & 5/37 (0.14) & 5/19 (0.26) & 38/38 (1) \\
    $\bm{Q_2}$ & 52/282 (0.18) & 5/38 (0.13) & 5/19 (0.26) & 37/37 (1) \\
    $\bm{Q_3}$ & 52/281 (0.18) & 5/37 (0.14) & 5/19 (0.26) & 37/37 (1) \\
    $\bm{Q_4}$ & 52/281 (0.18) & 5/37 (0.14) & 5/19 (0.26) & 37/37 (1) \\
    \midrule
    \multicolumn{5}{l}{\textbf{Scenario 2}} \\
    $\bm{Q_1}$ & 8/358 (0.02) & 5/10 (0.5) & 3/3 (1) & 4/4 (1) \\
    $\bm{Q_2}$ & 16/325 (0.05) & 5/26 (0.19) & 5/10 (0.5) & 14/14 (1) \\
    $\bm{Q_3}$ & 38/274 (0.14) & 5/48 (0.10) & 5/21 (0.24) & 32/32 (1) \\
    $\bm{Q_4}$ & 151/169 (0.89) & 5/67 (0.07) & 5/40 (0.13) & 99/99 (1) \\
    \midrule
    \multicolumn{5}{l}{\textbf{Scenario 3}} \\
    $\bm{Q_1}$ & 8/348 (0.02) & 5/15 (0.33) & 5/5 (1) & 7/7 (1) \\
    $\bm{Q_2}$ & 17/313 (0.05) & 5/31 (0.16) & 5/13 (0.38) & 18/18 (1) \\
    $\bm{Q_3}$ & 53/271 (0.20) & 5/45 (0.11) & 5/22 (0.23) & 37/37 (1) \\
    $\bm{Q_4}$ & 132/192 (0.69) & 5/60 (0.08) & 5/35 (0.14) & 88/88 (1) \\
    \bottomrule
\end{tabular*}
\end{table*}
\end{center}

Each simulation scenario was replicated $3{,}000$ times to evaluate the validity and efficiency of the design-analysis combinations summarized in Table~\ref{tab:simMethods}. Under SRS, the ACML estimator is equivalent to the standard maximum likelihood (ML) estimator, as the AC term in equation~(\ref{eq:acll}) does not involve any model parameters. For ODS and CSODS, ACML estimation incorporated the corresponding phase 2 sampling probabilities. When implementing the MI approach, we generated $C = 100$ candidate values for imputing $X$ and set $M = 100$ imputations. Under SRS + MI, the response model parameters in Step (1) of the MI procedure described in Section~\ref{sec:analysis} were estimated using standard maximum likelihood. For SMLE, we constructed B-spline basis for $Z_1$ using the \texttt{bs()} function in the \texttt{splines} package in R, specifying \texttt{df = 20} and \texttt{degree = 1}.

Table~\ref{tab:simResults_wide} evaluates the validity of statistical inferences across designs and analysis approaches for $\bbeta = (\beta_x, \beta_{z1}, \beta_{z2})$. Across scenarios, all design-estimation combinations yielded negligible bias, with average estimates across simulation replicates close to the true parameter values. The average standard error estimates (SEE) closely approximated the empirical standard errors (SE), and the resulting confidence intervals achieved approximately $95\%$ coverage. 

\begin{center}
\begin{table*}[!ht]%
\caption{Simulation results for estimating $\beta_x$, $\beta_{z1}$, and $\beta_{z2}$ across $3{,}000$ replicates. \label{tab:simResults_wide}}
\begin{tabular*}{\textwidth}{@{\extracolsep\fill}lllrrrrrr@{\extracolsep\fill}}
    \toprule
    \multicolumn{3}{l}{} & \multicolumn{2}{c}{\textbf{Scenario 1}} & \multicolumn{2}{c}{\textbf{Scenario 2}} & \multicolumn{2}{c}{\textbf{Scenario 3}} \\
    \cmidrule{4-5}\cmidrule{6-7}\cmidrule{8-9}
    & \textbf{Design} & \textbf{Estimation} & \textbf{EST [CP]} & \textbf{SE / SEE} & \textbf{EST [CP]} & \textbf{SE / SEE} & \textbf{EST [CP]} & \textbf{SE / SEE} \\
    \midrule
    \multirow[t]{10}{*}{$\bm{\beta}_x$} & \multirow[t]{3}{*}{\textbf{SRS}} & \textbf{ML} 
                      & $0.51$ [$0.95$] & $0.126$ / $0.125$ & $0.51$ [$0.94$] & $0.141$ / $0.137$ & $0.51$ [$0.95$] & $0.149$ / $0.145$ \\
    & & \textbf{MI}   & $0.51$ [$0.95$] & $0.126$ / $0.125$ & $0.51$ [$0.94$] & $0.141$ / $0.136$ & $0.51$ [$0.95$] & $0.149$ / $0.145$ \\
    & & \textbf{SMLE} & $0.52$ [$0.95$] & $0.130$ / $0.127$ & $0.52$ [$0.94$] & $0.145$ / $0.138$ & $0.52$ [$0.95$] & $0.152$ / $0.146$ \\
    & \multirow[t]{3}{*}{\textbf{ODS}} & \textbf{ACML} 
                      & $0.51$ [$0.95$] & $0.100$ / $0.100$ & $0.51$ [$0.95$] & $0.119$ / $0.115$ & $0.51$ [$0.94$] & $0.130$ / $0.127$ \\
    & & \textbf{MI}   & $0.50$ [$0.94$] & $0.096$ / $0.093$ & $0.50$ [$0.94$] & $0.116$ / $0.108$ & $0.51$ [$0.93$] & $0.127$ / $0.119$ \\
    & & \textbf{SMLE} & $0.51$ [$0.95$] & $0.100$ / $0.101$ & $0.51$ [$0.95$] & $0.119$ / $0.115$ & $0.51$ [$0.95$] & $0.124$ / $0.122$ \\
    & \multirow[t]{3}{*}{\textbf{CSODS}} & \textbf{ACML} 
                      & $0.51$ [$0.95$] & $0.099$ / $0.100$ & $0.51$ [$0.94$] & $0.103$ / $0.100$ & $0.51$ [$0.95$] & $0.120$ / $0.117$ \\
    & & \textbf{MI}   & $0.50$ [$0.94$] & $0.096$ / $0.093$ & $0.50$ [$0.94$] & $0.097$ / $0.093$ & $0.50$ [$0.94$] & $0.118$ / $0.112$ \\
    & & \textbf{SMLE} & $0.51$ [$0.95$] & $0.100$ / $0.101$ & $0.51$ [$0.94$] & $0.102$ / $0.096$ & $0.50$ [$0.94$] & $0.115$ / $0.109$\\
    & \textbf{RDS} & \textbf{SMLE} 
                      & $0.50$ [$0.95$] & $0.108$ / $0.106$ & $0.51$ [$0.94$] & $0.111$ / $0.107$ & $0.51$ [$0.94$] & $0.093$ / $0.089$ \\
    \midrule
    \multirow[t]{10}{*}{$\bm{\beta}_{z1}$} & \multirow[t]{3}{*}{\textbf{SRS}} & \textbf{ML} 
                      & $-0.10$ [$0.95$] & $0.121$ / $0.120$ & $-1.53$ [$0.95$] & $0.176$ / $0.171$ & $-1.53$ [$0.95$] & $0.185$ / $0.181$ \\
    & & \textbf{MI}   & $-0.10$ [$0.95$] & $0.069$ / $0.069$ & $-1.51$ [$0.94$] & $0.104$ / $0.102$ & $-1.51$ [$0.95$] & $0.108$ / $0.108$ \\
    & & \textbf{SMLE} & $-0.10$ [$0.95$] & $0.070$ / $0.069$ & $-1.52$ [$0.95$] & $0.106$ / $0.106$ & $-1.52$ [$0.95$] & $0.111$ / $0.111$ \\
    & \multirow[t]{3}{*}{\textbf{ODS}} & \textbf{ACML} 
                      & $-0.10$ [$0.95$] & $0.097$ / $0.096$ & $-1.52$ [$0.95$] & $0.147$ / $0.146$ & $-1.52$ [$0.95$] & $0.160$ / $0.160$ \\
    & & \textbf{MI}   & $-0.10$ [$0.95$] & $0.068$ / $0.068$ & $-1.51$ [$0.94$] & $0.102$ / $0.098$ & $-1.51$ [$0.95$] & $0.105$ / $0.104$ \\
    & & \textbf{SMLE} & $-0.10$ [$0.95$] & $0.067$ / $0.067$ & $-1.52$ [$0.95$] & $0.104$ / $0.103$ & $-1.52$ [$0.96$] & $0.105$ / $0.107$ \\
    & \multirow[t]{3}{*}{\textbf{CSODS}} & \textbf{ACML}
                      & $-0.10$ [$0.95$] & $0.095$ / $0.096$ & $-1.51$ [$0.95$] & $0.128$ / $0.126$ & $-1.51$ [$0.95$] & $0.151$ / $0.148$ \\
    & & \textbf{MI}   & $-0.10$ [$0.96$] & $0.067$ / $0.068$ & $-1.50$ [$0.95$] & $0.097$ / $0.097$ & $-1.50$ [$0.95$] & $0.104$ / $0.104$ \\
    & & \textbf{SMLE} & $-0.10$ [$0.95$] & $0.066$ / $0.067$ & $-1.51$ [$0.94$] & $0.094$ / $0.092$ & $-1.51$ [$0.95$] & $0.099$ / $0.098$\\
    & \textbf{RDS} & \textbf{SMLE}
                      & $-0.10$ [$0.94$] & $0.068$ / $0.068$ & $-1.51$ [$0.95$] & $0.091$ / $0.090$ & $-1.51$ [$0.95$] & $0.094$ / $0.094$ \\
    \midrule
    \multirow[t]{10}{*}{$\bm{\beta}_{z2}$} & \multirow[t]{3}{*}{\textbf{SRS}} & \textbf{ML} 
                      & $-0.10$ [$0.95$] & $0.121$ / $0.118$ & $-0.10$ [$0.95$] & $0.131$ / $0.129$ & $-1.54$ [$0.95$] & $0.182$ / $0.180$ \\
    & & \textbf{MI}   & $-0.10$ [$0.95$] & $0.065$ / $0.064$ & $-0.10$ [$0.95$] & $0.071$ / $0.070$ & $-1.51$ [$0.95$] & $0.099$ / $0.099$ \\
    & & \textbf{SMLE} & $-0.10$ [$0.95$] & $0.062$ / $0.062$ & $-0.10$ [$0.95$] & $0.068$ / $0.068$ & $-1.51$ [$0.96$] & $0.098$ / $0.099$ \\
    & \multirow[t]{3}{*}{\textbf{ODS}} & \textbf{ACML} 
                      & $-0.10$ [$0.96$] & $0.092$ / $0.094$ & $-0.10$ [$0.94$] & $0.112$ / $0.109$ & $-1.53$ [$0.95$] & $0.162$ / $0.158$ \\
    & & \textbf{MI}   & $-0.10$ [$0.95$] & $0.065$ / $0.065$ & $-0.10$ [$0.95$] & $0.071$ / $0.071$ & $-1.51$ [$0.95$] & $0.099$ / $0.098$ \\
    & & \textbf{SMLE} & $-0.10$ [$0.95$] & $0.062$ / $0.062$ & $-0.10$ [$0.95$] & $0.068$ / $0.067$ & $-1.51$ [$0.95$] & $0.096$ / $0.097$ \\
    & \multirow[t]{3}{*}{\textbf{CSODS}} & \textbf{ACML} 
                      & $-0.10$ [$0.95$] & $0.094$ / $0.094$ & $-0.10$ [$0.95$] & $0.095$ / $0.093$ & $-1.52$ [$0.95$] & $0.153$ / $0.150$ \\
    & & \textbf{MI}   & $-0.10$ [$0.95$] & $0.066$ / $0.065$ & $-0.10$ [$0.95$] & $0.076$ / $0.075$ & $-1.51$ [$0.95$] & $0.101$ / $0.101$ \\
    & & \textbf{SMLE} & $-0.10$ [$0.95$] & $0.062$ / $0.062$ & $-0.10$ [$0.95$] & $0.068$ / $0.067$ & $-1.51$ [$0.95$] & $0.096$ / $0.094$\\
    & \textbf{RDS} & \textbf{SMLE} 
                      & $-0.10$ [$0.95$] & $0.061$ / $0.061$ & $-0.10$ [$0.95$] & $0.068$ / $0.067$ & $-1.51$ [$0.95$] & $0.093$ / $0.092$ \\
    \bottomrule
\end{tabular*}
\begin{tablenotes}
\item {\it Abbreviations:} 
EST = average estimate;
CP = coverage probability of the 95\% confidence interval;
SE = empirical standard error;
SEE = average standard error estimate;
SRS = simple random sampling; 
ODS = outcome-dependent sampling; 
CSODS = covariate-stratified outcome-dependent sampling; 
RDS = residual-dependent sampling; 
ML = maximum likelihood estimation; 
ACML = ascertainment-corrected maximum likelihood estimation; 
MI = multiple imputation; 
SMLE = sieve maximum likelihood estimation.
\end{tablenotes}
\end{table*}
\end{center}

\begin{figure}[ht]
\centerline{\includegraphics[width=0.93\textwidth]{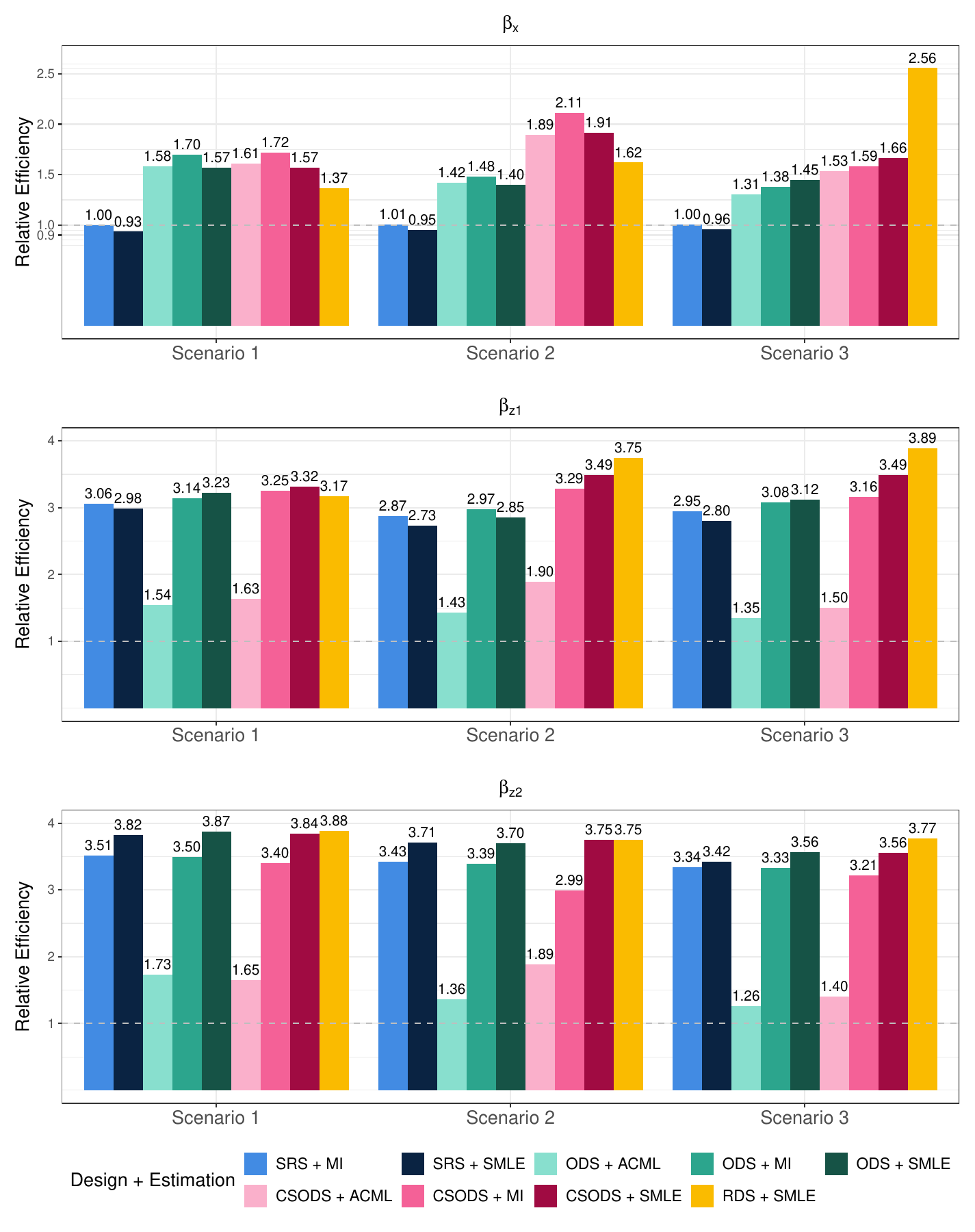}}
\caption{Relative efficiency of the nine design-estimation combinations across three simulation scenarios, using SRS + ML as the reference within each scenario. Relative efficiency is calculated as the empirical variance of SRS + ML divided by that of each combination. \label{fig:simRE}}
\begin{flushleft}
\footnotesize
\textit{Abbreviations:} 
SRS = simple random sampling; 
ODS = outcome-dependent sampling; 
CSODS = covariate-stratified outcome-dependent sampling; 
RDS = residual-dependent sampling; 
ML = maximum likelihood estimation; 
ACML = ascertainment-corrected maximum likelihood estimation; 
MI = multiple imputation; 
SMLE = sieve maximum likelihood estimation.
\end{flushleft}
\end{figure}

Figure~\ref{fig:simRE} summarizes the relative efficiency (RE) of nine design-estimation combinations across three simulation scenarios, using SRS + ML as the reference. In Scenario 1, where both $Z_1$ and $Z_2$ were weakly associated with $Y$, all three proposed designs significantly improved efficiency for the exposure effect $\beta_x$ relative to SRS under the same estimation procedure (top row). ODS and CSODS achieved REs of $1.58$ and $1.61$ with ACML, and $1.70$ and $1.72$ with MI. RDS was evaluated only with SMLE, where its efficiency gain ($1.37/0.93=1.46$) was lower than that of ODS ($1.57/0.93=1.68$) and CSODS ($1.57/0.93=1.68$). This reduced efficiency of RDS may be due to residuals capturing noise rather than informative variability in $Y \mid Z_1, Z_2$ when the covariates in the working model are only weakly associated with the outcome. Across estimation methods, MI improved efficiency over ACML by $7\%$ under both ODS (RE $1.70$ vs. $1.58$) and CSODS ($1.72$ vs. $1.61$). 

When one or more covariates were strongly associated with $Y$ (Scenarios 2 and 3), efficiency gains for the exposure effect $\beta_x$ became more pronounced, with the covariate-informed designs outperforming both SRS and ODS. In Scenario 2 ($\beta_{z1} = -1.5$), CSODS improved efficiency by $43\%$ relative to ODS under MI (RE $2.11$ vs. $1.48$). Under SMLE, both CSODS and RDS outperformed ODS, with CSODS + SMLE achieving a 37\% efficiency gain (RE $1.91$ vs. $1.40$) and RDS + SMLE achieving a 16\% gain (RE $1.62$ vs. $1.40$). In Scenario 3, where both $Z_1$ and $Z_2$ were strongly associated with $Y$ ($\beta_{z1} = \beta_{z2} = -1.5$), RDS achieved the largest efficiency gains. RDS + SMLE yielded a RE of $2.56$, substantially outperforming CSODS + SMLE (RE $1.66$) by $54\%$, ODS + SMLE (RE $1.45$) by $77\%$, and SRS + SMLE (RE $0.96$) by $167\%$. 

For the covariate effects $\beta_{z1}$ and $\beta_{z2}$ (middle and bottom rows), efficiency gains from proposed sampling designs were modest under the complete-case ACML analysis, whereas full-data estimation procedures (MI and SMLE) were far more efficient. Overall, sampling design had a greater impact on the efficiency of the exposure effect $\beta_x$, whereas the estimation procedure played a larger role for covariate effects $(\beta_{z1}, \beta_{z2})$ that are fully observed in phase 1. This has been observed in other work \cite{schildcrout2015, digravio2024}.

\section{Applications to the CLOVERS} \label{sec:clovers}

The CLOVERS trial was a multi-center randomized clinical trial involving patients with early sepsis \cite{clovers2023}. Its primary objective was to evaluate the association between a restrictive versus liberal fluid strategy and mortality prior to discharge home by day 90. In addition to the primary endpoint, the trial collected secondary data that allowed construction of a four-level ordinal measure of daily clinical status recorded over 28 days (coded as 1 = discharged, 2 = hospitalized but not in the ICU with invasive support, 3 = hospitalized in the ICU with invasive support, and 4 = death). Biospecimens were also collected and stored for retrospective biomarker analysis. Interleukin-6 (IL-6), an inflammatory biomarker, was measured as part of the trial protocol for 1,371 of the 1,563 participants.

In this section, we considered a secondary analysis to evaluate the association between IL-6 levels and day 14 clinical status. We restricted attention to the 1,351 participants with complete data on IL-6, baseline covariates, and day 14 clinical status, which we treated as the phase 1 cohort. To mimic resource constraints commonly encountered in biomarker studies, we assumed that IL-6 measurements could be obtained for only 400 phase 2 participants. Each sampling design (SRS, ODS, CSODS, and RDS) was used to select a different subset of 400 participants from the phase 1 cohort, and the performance of each design was evaluated by comparing results from these subsets to those obtained using data from all 1,351 participants. For the SRS, ODS, and CSODS designs, phase 2 sampling was repeated 100 times, and regression coefficients and their variance-covariance matrices were averaged across replicates. Because the RDS design is deterministic---selecting 200 participants from each tail of the residual distribution---it was evaluated using a single replicate.

For each sampled dataset, we fit a proportional odds model for day 14 clinical status that included log-transformed IL-6, treatment assignment, age, sex, race/ethnicity, Sequential Organ Failure Assessment (SOFA) score, and history of kidney disease and heart failure. Restricted cubic splines were used to allow nonlinear associations for IL-6, age, and SOFA score. Because nonlinear modeling of IL-6 precludes direct comparison of individual coefficient estimates and uncertainty across designs, we compared methods using IL-6 association plots with 95\% point-wise confidence intervals over the 2.5th to 97.5th percentiles of observed IL-6 distribution. In addition, we summarized uncertainty in the estimated IL-6 association using two measures: (i) estimated D-efficiency of the IL-6 parameters, defined as $\left|\widehat{\text{Cov}}(\hat{\bm{\beta}}_{IL-6})^{-1}\right|^{1/p}$, where $\hat{\bm\beta}_{IL-6}$ denotes the vector of regression coefficients associated with IL-6, $\widehat{\text{Cov}}(\hat{\bm{\beta}}_{IL-6})$ is its estimated variance-covariance matrix, and $p$ is the number of IL-6 spline parameters; and (ii) the integrated area of the IL-6 association plots, defined as the numerical integral of the confidence interval width over its IL-6 range.

\begin{figure}[ht]
\centerline{\includegraphics[width=0.5\textwidth]{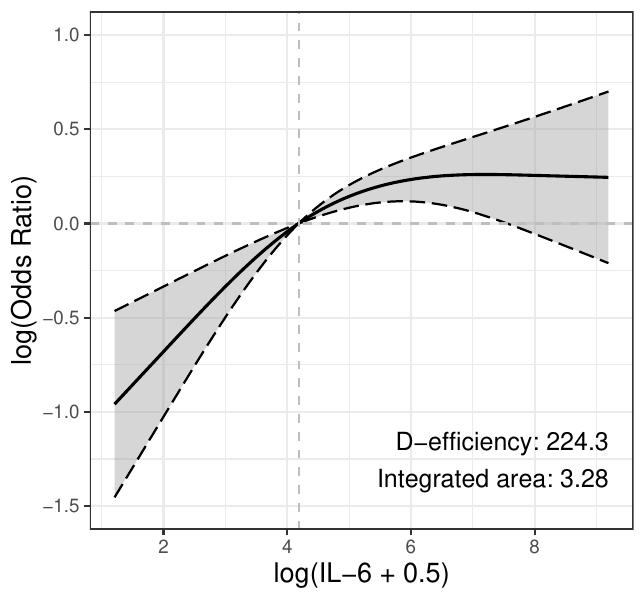}}
\caption{Estimated association between interleukin-6 (IL-6) and Day 14 clinical status in the full CLOVERS trial cohort. The solid line represents the estimated log-odds ratio as a function of log-transformed IL-6. The shaded region shows point-wise 95\% confidence intervals. Efficiency metrics include estimated D-efficiency and the integrated area of the shaded region. \label{fig:clovers_fullcohort}}
\end{figure}

\begin{figure}[ht]
\centerline{\includegraphics[width=0.88\textwidth]{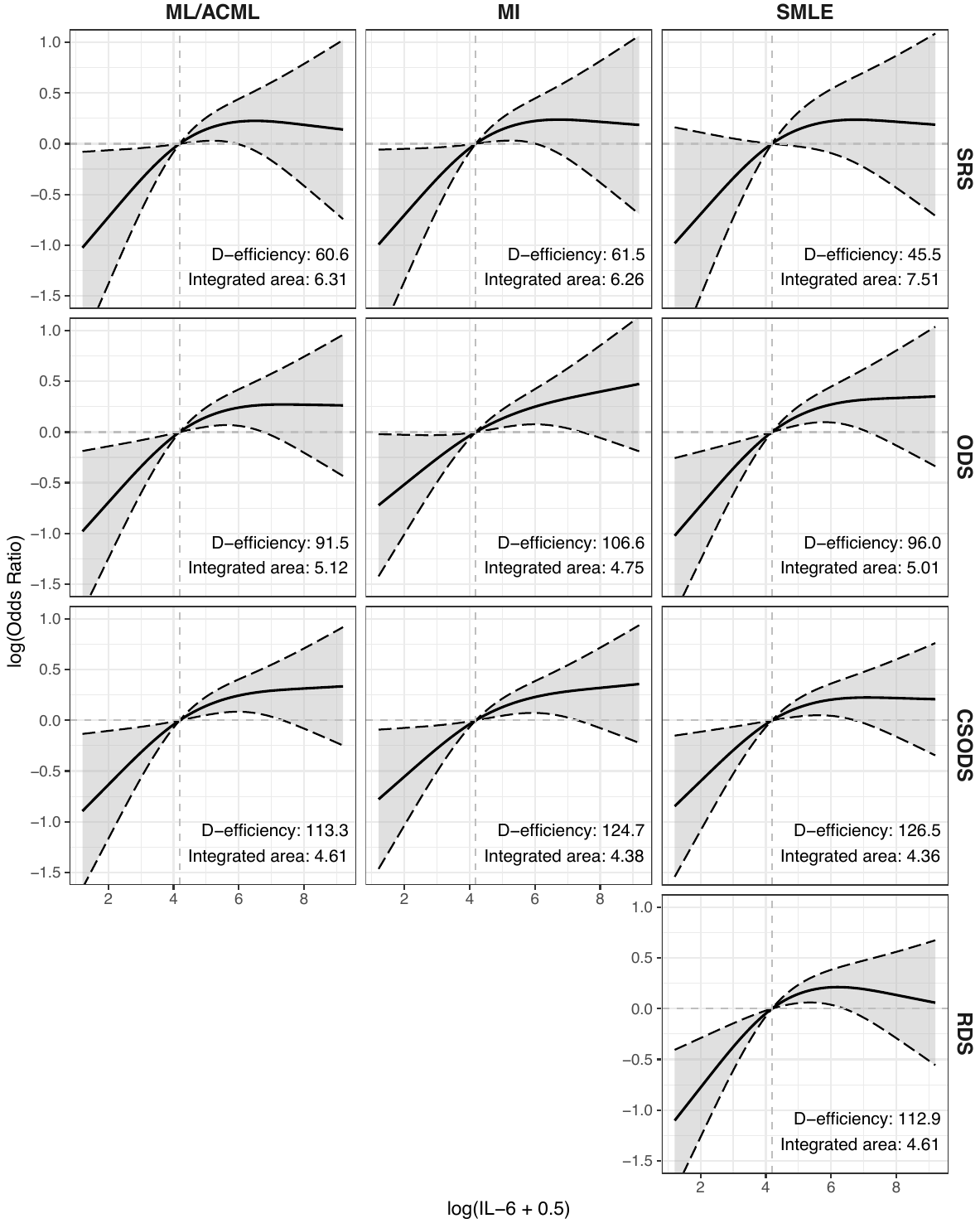}}
\caption{Estimated association between interleukin-6 (IL-6) and Day 14 clinical status in phase-two samples from the CLOVERS cohort evaluated across sampling designs and analysis procedures using a plasmode simulation. The solid line represents the estimated log-odds ratio as a function of log-transformed IL-6. The shaded region shows point-wise 95\% confidence intervals. Efficiency metrics include estimated D-efficiency and the integrated area of the shaded region. \label{fig:clovers_phasetwo}}
\begin{flushleft}
\footnotesize
\textit{Abbreviations:} 
SRS = simple random sampling; 
ODS = outcome-dependent sampling; 
CSODS = covariate-stratified outcome-dependent sampling; 
RDS = residual-dependent sampling; 
ML = maximum likelihood estimation; 
ACML = ascertainment-corrected maximum likelihood estimation; 
MI = multiple imputation; 
SMLE = sieve maximum likelihood estimation.
\end{flushleft}
\end{figure}

Figure~\ref{fig:clovers_fullcohort} illustrates the estimated association between IL-6 concentration and the outcome using the full cohort of 1,351 participants. The estimated association increases monotonically with IL-6 concentration, with steeper changes at lower IL-6 values and gradual flattening at higher values. This full-data estimate serves as the benchmark for evaluating the design-analysis strategies under two-phase sampling. Figure~\ref{fig:clovers_phasetwo} shows the estimated IL-6 association curves for all design-analysis combinations listed in Table~\ref{tab:simMethods}. Across methods, the overall shape of the association was well preserved despite using only 400 IL-6 measurements at phase 2, indicating that all approaches appear to recover the underlying trend present in the full cohort. 

Although the overall shape was similar across methods, the choices of phase 2 sampling design and estimation procedure had a notable impact on uncertainty. Outcome-informed phase 2 sampling strategies consistently yielded lower uncertainty than SRS, reflected by both larger estimates of D-efficiency and smaller integrated areas. For instance, under SMLE, estimated D-efficiency increased from 45.5 for SRS to 96.0 for ODS, 126.5 for CSODS, and 112.9 for RDS, while the integrated area decreased from 7.51 for SRS to 5.01, 4.36, and 4.61 for ODS, CSODS, and RDS, respectively. Comparing across analysis procedures within each design, under ODS and CSODS, SMLE and MI outperformed ML/ACML in terms of both metrics. Under CSODS, for example, estimated D-efficiency increased from 113.3 under ACML to 124.7 under MI and 126.5 under SMLE. 

The empirical findings in CLOVERS were consistent with the simulation results, particularly those from Scenario 2 in which a single covariate was strongly associated with the outcome. In this setting, SOFA score played a comparable role to the strongly predictive covariate in the simulations, and the covariate-informed designs---CSODS and ODS---achieved lower uncertainty than ODS and SRS. The diminished performance of RDS can be attributed to its dependence on residuals from a working model that included all baseline covariates, some of which were only weakly associated with the outcome, yielding residuals that reflect noise in addition to informative variability.

\section{Discussion} \label{sec:discuss}

Motivated by the increasing use of ordinal endpoints in biomedical research, we developed a novel class of two-phase study designs and inference procedures for ordinal outcomes under the proportional odds model. These methods address a common challenge in epidemiological research where investigators aim to estimate exposure-outcome associations but budgetary constraints limit measurement of expensive exposures to only a subset of participants. Our comprehensive framework provides investigators with multiple design and analysis options that can be tailored to their scientific objectives, available baseline data, and computational considerations.

We proposed three outcome-informed phase 2 sampling designs---ODS, CSODS, and RDS---that leverage phase 1 data to enrich phase 2 selection with informative subjects. The success of these designs depends on how effectively they capture variability in the exposure-outcome association. ODS improves efficiency over SRS by oversampling extreme outcome categories but may be suboptimal when baseline covariates explain substantial outcome variability. In such settings, designs that incorporate covariate information are recommended. As shown in Section~\ref{sec:sim}, when covariates were weakly associated with the outcome (Scenario 1), all three designs achieved comparable gains over SRS. When one or more covariates were strongly associated with the outcome in Scenarios 2-3, CSODS and RDS outperformed ODS, with RDS providing the largest gains when multiple strong predictors were present. The efficiency of RDS depended on the informativeness of the working model used to compute PSRs; when residuals were based on weakly predictive covariates, gains were modest. Overall, CSODS is an effective choice when a key covariate is available, whereas RDS can provide additional efficiency gains in settings with multiple such covariates.

We further examined three estimation procedures that provide valid inference under biased phase 2 sampling: ACML, MI, and SMLE. ACML is a complete-case approach that is useful when, for those not selected into the phase 2 sample, phase 1 covariates are unavailable or cannot be accessed due to data sharing restrictions. When phase 1 covariates are available, full-data approaches such as MI and SMLE may be preferable because they recover information from unsampled subjects. In our simulations, MI consistently improved efficiency over ACML under ODS and CSODS when the imputation model was correctly specified. SMLE offers a semiparametric alternative that avoids specification of an exposure model by estimating the exposure distribution nonparametrically. In settings where multiple covariates are correlated with the exposure, careful selection of the number of B-spline basis functions is required to balance model flexibility and computational stability. The current MI approach is not compatible with RDS because it relies on ACML to fit the response model among sampled subjects. However, a direct conditional exposure model (Schildcrout et al., 2015) could be extended to incorporate RDS and represents a promising direction for future work. Another important avenue for future research involves settings where the outcome or exposure model is misspecified. It is known that design-based analysis approaches offer robustness to several misspecified model settings. It will therefore be important to examine the performance of the likelihood-based analysis approaches discussed here under model misspecification.

\bmsubsection*{Acknowledgments}

This research was supported by the National Institute of Health grant R01HL094786 and R01AI131771. This work was conducted in part using the resources of the Advanced Computing Center for Research and Education at Vanderbilt University, Nashville, TN.

\bibliography{wileyNJD-AMA}

\appendix

\bmsection{EM Algorithm for the SMLE}

The second term in equation~(\ref{eq:obs_ll2_sml}) involves a logarithm of a summation, which complicates direct maximization. Following Tao et al. (2017), we introduce a latent variable $U_i$ for subjects with $S_i = 0$, defined such that:
\begin{itemize}
    \item $U_i \sim P(U_i = l/s_n \mid \bZ_i) = B_l^q(\bZ_{1i})$ for $l=1, \dots, s_n$,
    \item $P(X_i = x_v \mid \bZ_i, U_i = l/s_n) = P(X_i = x_v \mid U_i = l/s_n) = p_{vl}$, and
    \item $P(Y_i \mid X_i, \bZ_i, U_i) = P(Y_i \mid X_i, \bZ_i)$.
\end{itemize}
Under this specification, for $S_i = 0$, the exposure distribution can be expressed as:
\begin{equation} \label{eq:GwithU}
    H(X_i = x_v \mid \bZ_i) 
    = \sum_{l=1}^{s_n} B_l^q(\bZ_{1i})\, p_{vl}
    = \sum_{l=1}^{s_n} P(X_i = x_v \mid U_i = l/s_n)\, P(U_i = l/s_n \mid \bZ_i).
\end{equation}
Because $U_i$ is defined through the B-spline basis functions of $\bZ_{1i}$, its distribution given $\bZ_i$ introduces no additional parameters. The latent variable representation transforms the log-sum term into a tractable complete-data form, yielding the following complete-data log-likelihood:
\begin{equation} \label{eq:complete_ll_smle}
    \begin{split}
        l_N^f(\btheta, \{p_{vl}\}) \propto 
        &\sum_{i=1}^N S_i\left\{\log P (Y_i \mid X_i, \bZ_i; \btheta) + \sum_{v=1}^d I(X_i = x_v) \sum_{l=1}^{s_n} B_l^q(\bZ_{1i})\log p_{vl}\right\} \\
        & + \sum_{i=1}^N (1 - S_i)\Biggl\{\sum_{v=1}^d  I(X_i = x_v) \log P (Y_i \mid x_v, \bZ_i; \btheta)
            + \sum_{v=1}^d \sum_{l=1}^{s_n}I(X_i = x_v, U_i = l/s_n)\, \log p_{vl}\Biggr\}.
    \end{split}
\end{equation}

We develop an EM algorithm to maximize expression~(\ref{eq:obs_ll2_sml}). The complete data consist of $(Y_i, X_i, \bZ_i, U_i)$, where $X_i$ and $U_i$ are considered missing for subjects with $S_i = 0$. The EM algorithm proceeds as follows:
\begin{enumerate}
    \item Specify initial values: 
    $$
    \hat{\btheta}^{(0)} = \mathbf{0}\text{ and }
    \hat{p}_{vl}^{(0)} = \frac
    {\sum_{i=1}^N S_i\, I(X_i = x_v)\, B_l^q(\bZ_{1i})}
    {\sum_{i=1}^N \sum_{v=1}^d S_i\, I(X_i = x_v)\, B_l^q(\bZ_{1i})}\text{ for }v = 1,\ldots, d, \text{ and } l = 1,\ldots, s_n.
    $$
    \item At the $b$th iteration,
    \begin{enumerate}[a.]
        \item \textbf{E-step:} For subjects with $S_i = 0$, compute the conditional expectations:
        \begin{equation*}
            \begin{split}
                \hat{\psi}_{ivl}^{(b)}
                 = E\left[I\left\{ (X_i, U_i) = (x_v, l/s_n)\right\} \mid Y_i, \bZ_i; \hat{\btheta}^{(b-1)}, \hat{p}_{vl}^{(b-1)}\right]
                 &= 
                \frac{P(Y_i \mid x_v, \bZ_i; \hat{\btheta}^{(b-1)})\, B_l^q(\bZ_{1i})\, \hat{p}_{vl}^{(b-1)}}
                {\sum_{v'=1}^{d} \sum_{l'=1}^{s_n} P(Y_i \mid x_{v'}, \bZ_i; \hat{\btheta}^{(b-1)})\, B_{l'}^q(\bZ_{1i})\, \hat{p}_{v'l'}^{(b-1)}}\\ 
                \text{ and }
                \hat{q}_{iv}^{(b)} = E\left[I(X_i=x_v) \mid Y_i, \bZ_i; \hat{\btheta}^{(b-1)}, \hat{p}_{vl}^{(b-1)}\right]
                &= \sum_{l=1}^{s_n} \hat{\psi}_{ivl}^{(b)}
            \end{split}
        \end{equation*}
        \item \textbf{M-step:} Update the parameter estimates by maximizing the expected log-likelihood:
        \begin{enumerate}[i.]
            \item Update $\btheta$ by maximizing the weighted log-likelihood:
            $$
            \hat{\btheta}^{(b)} = \underset{\btheta}{\text{argmax}} \sum_{i=1}^N S_i\, \log P (Y_i \mid X_i, \bZ_i; \btheta) + 
            \sum_{i=1}^N (1 - S_i)\, \sum_{v=1}^d  \hat{q}_{iv}^{(b)}\, \log P (Y_i \mid x_k, \bZ_i; \btheta)
            $$
            \item Update $p_{vl}$ by maximizing:
            $$
            \sum_{i=1}^N S_i\, \sum_{v=1}^d \sum_{l=1}^{s_n} I(X_i = x_k)\, B_l^q(\bZ_{1i})\, \log p_{vl} + 
            \sum_{i=1}^N (1 - S_i)\, \sum_{v=1}^d \sum_{l=1}^{s_n} \hat{\psi}_{ivl}^{(b)}\, \log p_{vl}
            $$
            such that
            $$
            p_{vl}^{(b)} = \frac
            {\sum_{i=1}^N \left\{ S_i\, I(X_i = x_k)\, B_l^q (\bZ_{1i}) + (1 - S_i)\, \hat{\psi}_{ivl}^{(b)} \right\}}
            {\sum_{i=1}^N \sum_{v=1}^d \left\{ S_i\, I(X_i = x_k)\, B_l^q(\bZ_{1i}) + (1 - S_i)\, \hat{\psi}_{ivl}^{(b)} \right\}}
            $$
        \end{enumerate}
    \end{enumerate}
    \item Repeat the E-step and M-step until convergence.
\end{enumerate}
Note that existing software packages can be used to obtain the parameter estimates $\hat{\btheta}^{(b)}$ that maximize the weighted log-likelihood in the M-step of iteration $b$. 

\end{document}